\documentstyle[pra,aps,twocolumn,epsfig]{revtex}

\begin{document}

\title{Bloch-Like Quantum Multiple Reflections of Atoms}

\author{Luis Santos and Luis Roso}

\address{Departamento de F\'\i sica Aplicada \\
Universidad de Salamanca, 37008 Salamanca, Spain}

\maketitle

\begin{abstract}
We show that under certain circumstances an atom can follow an oscillatory motion in a periodic
laser profile with a Gaussian envelope. These oscillations can be well explained by using a model
of energetically forbidden spatial regions. The similarities and differences with Bloch
oscillations are discussed. We demonstrate that the effect exists not only for repulsive but also
for attractive potentials, i.e. quantum multiple reflections are also possible.
\end{abstract}


\section{Introduction}
\label{sec:intro}

During the last years the development of different and powerful laser cooling techniques
\cite{Chu98} has lead to the experimental observation of several striking
matter--wave phenomena for cold atoms. These phenomena resemble the usual effects observed in
standard optics, and therefore they have been the subject of study of a fast--developing research
field properly called {\em atom optics}. Some atom optics examples could be the diffraction of atoms
\cite{Keith88}, interferometers\cite{Carnal91} and lenses \cite{Carnal91b} for atoms, and very
recently even the atom optics equivalent of a laser source, or atom--laser \cite{Andrews97}. Atomic
mirrors have also been constructed using evanescent laser fields \cite{Balykin87} or, more
recently, with magnetic fields \cite{Roach95}. Both mirrors, although based on different physical
effects, share the same physical background, i.e. the atom feels a repulsive potential, and if its
initial velocity is sufficiently small, it reaches a turning--point, being reflected. Note that
after the turning--point the atomic density of probability follows an exponential decay, resembling
the case of an electric field impinged onto a metallic surface. Therefore, these atomic mirrors
can be considered the atom optics counterpart of metallic mirrors in light optics.

On the other hand, in the last few years, the behavior of a cold atom in periodic laser potentials
(light lattices) has aroused great interest, in particular the resemblance of this physical
situation and Solid--State physics \cite{Wilkens91}. In this sense, Bragg scattering has been
analysed in the context of atomic waves \cite{Martin88}, and so has the use of such effect to
construct atomic beam--splitters and interferometers \cite{Rasel95}. Bragg reflection is just a
particular case of a more general situation, i.e. the so--called Photonic--Band--Gap--Structures
(or PBGS) \cite{Yablonovitch87}. In dielectric periodic structures some electromagnetic waves
cannot propagate (basically because their corresponding energies are within an energetic gap
produced by the periodic structure), and therefore are reflected. In previous papers
\cite{Santos97} we have proposed a laser arrangement that acts as an Atomic--Band--Gap--Structure,
in the sense that it resembles a PBGS but for atomic waves instead of electromagnetic ones. The
atoms are reflected if their incoming kinetic energy lies within a gap produced by the laser
periodicity. This effect produces a band--like momentum spectrum of the reflected atoms. We
have shown that the atoms can be reflected not only by repulsive laser potentials, but also by
attractive ones, allowing the atom optics equivalent of a multilayer dielectric mirror in light
optics.

Among the different Solid--State--like phenomena recently reported, is of special interest the
analysis of the behavior of cold atoms accelerated in a periodic potential. In this sense, the
well--known Bloch Oscillations (BO), and their stationary counterpart, i.e. the Wannier--Stark (WS)
ladders, have been experimentally observed in the context of atom optics
\cite{Dahan96}.

In the present paper we demonstrate that an atomic beam dropped onto a laser potential as that of
\cite{Santos97} can undergo under certain conditions multiple oscillations inside the laser region.
We show that this effect can be physically explained using an image of energetically forbidden
spatial regions, what we call spatial gaps. In particular the effect is produced due to a
combination of partial Landau--Zener tunneling through sufficiently narrow spatial gaps, and a
partial reflection on them. In this sense the laser arrangement behaves as an atomic--wave
Fabry-P\'erot interferometer \cite{Santos97b}. We have also analysed the similarities and
differences between these oscillations and Bloch oscillations. Due to these similarities and
differences we have called the effect Bloch--like oscillations. We show that the effect can be
easily observed analysing either the reflected momentum spectrum (where Wannier--Stark--like
resonances appear) or the temporal evolution of an atomic wavepacket. We prove that the multiple
reflections appear also for attractive potentials, i.e. quantum multiple reflections are also
possible.

The scheme of the paper is as follows. In Sec.\ \ref{sec:Theo} we briefly review the theoretical
model already presented in \cite{Santos97}. Sec.\ \ref{sec:Spgaps} is devoted to the development of
the spatial gaps image. Sec.\ \ref{sec:Fabry} uses the spatial gaps image to explain the appearance
of resonancies in the momentum spectrum of the reflected atoms, and compare the effect with a
Fabry--P\'erot interferometer. Sec.\ \ref{sec:Bows} discuss the similarities and differences
between the presented oscillations and Bloch oscillations. In Sec.\ \ref{sec:Bloch} we present the
temporal evolution of an atomic wavepacket for the case of a repulsive potential, whereas in Sec.\
\ref{sec:Qmr} the case of quantum multiple reflections is considered. We finalize in Sec.\
\ref{sec:Conclu} with some conclusions.

\section{Theoretical Model}
\label{sec:Theo}

\begin{figure}[ht]
\begin{center}\
\hspace{0mm}
\psfig{file=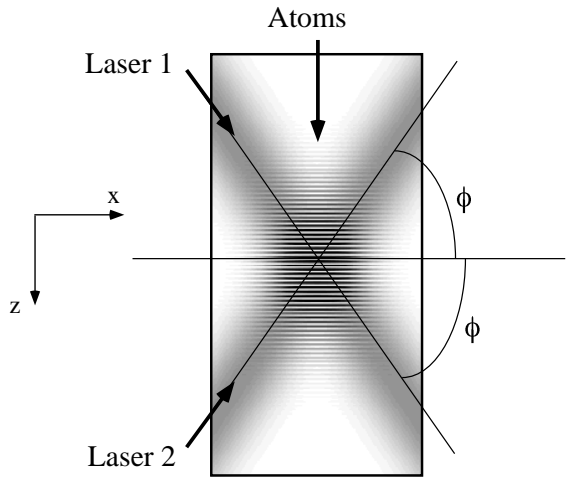,width=7.5cm}\\[0.3cm]
\caption{Scheme of the considered laser arrangement. Two Gaussian lasers propagate, forming
and angle $\phi$ and $-\phi$ with the $x$ axis. The atoms are dropped from a MOT onto the
interference region of both lasers, where a periodic profile is formed.}
\label{fig:1}
\end{center}
\end{figure}

In this section we briefly discuss the theoretical model we use in the paper. For a more detailed
discussion see \cite{Santos97}. Let us consider the same laser arrangement as that of
\cite{Santos97}, formed by the interference of two laser beams (Fig.\ \ref{fig:1}) of the same
intensity, polarisation, and with the same Gaussian profile, but with respective wave vectors: $\vec
k_{1}=k(\cos\phi\vec u_{x}+\sin\phi\vec u_{z})$ and $\vec k_{2}=k(\cos\phi\vec
u_{x}-\sin\phi\vec u_{z})$. In the zone very close to $x=0$ the electric field, which is
assumed to be linearly polarised in the $y$-direction, can be written in the form:
\begin{eqnarray}
\vec E&\simeq& E_{0}\vec e_{y}
e^{\frac{-(z-z_{c})^{2}\cos^{2}\phi}{d^{2}}}\cos(k(z-z_{c})\sin\phi) \nonumber \\
&\times&\cos(kx\cos\phi-\omega t),
\label{Efield}
\end{eqnarray}
where $E_{0}/2$ is the amplitude of each laser, $z_{c}$ is the center of the laser region, and $d$
is the halfwidth of the Gaussians. Let us assume that the laser frequency $\omega$ is
quasi--resonant with some atomic transition between the ground state and an excited state (of
energies $\hbar\omega_{g}$ and $\hbar\omega_{e}$ , respectively), and therefore we can treat the
atom as a two-level system. In this paper, we assume a large internal detuning
($\Delta=\omega-(\omega_{e}-\omega_{g})$)), in such a way that the adiabatic approximation will be
valid \cite{Deutschmann93}. In order to obtain a scalar Schr\"odinger equation to describe the
atomic interaction with the laser, we follow the standard formalism developed in
\cite{Deutschmann93}, except for the inclusion of the gravitational field. The scalar equation
takes the form:
\begin{equation}
\frac{-\hbar^{2}}{2M}\frac{d^{2}}{dz^{2}}\psi(z)=\left [\frac{q^{2}}{2M}-V(z)\right ]\psi(z),
\label{Scheq}
\end{equation}
where
\begin{equation}
V(z)=-Mgz-\frac{1}{2}\hbar\Delta\mp\frac{1}{2}\hbar\sqrt{\Delta^{2}+4\Omega(z)^{2}},
\label{Vz}
\end{equation}
with $\Omega(z)=\Omega_{0}\exp\left [-\cos^{2}\phi(z-z_{c})^{2}/d^{2}\right ]\cos
(k(z-z_{c})\sin\phi)$. The sign in Eq.\ (\ref{Vz}) depends on the detuning: the "--" sign
corresponds to $\Delta<0$, and the "+" corresponds to $\Delta>0$. The coupling is given by
$\Omega_{0}=-\mu E_{0}/2\hbar$, which is the Rabi frequency associated with each laser, where
$\mu=\langle\vec e_{y}\vec\mu\rangle$, with $\vec\mu$ the transition dipole. $q$ is the
$z$-momentum in $z=0$, sufficiently far away from $z_{c}$ to consider that the field in $z=0$ is
negligible. To make the model as realistic as possible one could include the diffuse scattering
resulting from spontaneous emission by adopting the lossy vector Schr\"odinger equation method of
\cite{Zhang94}, i.e. $\Delta\rightarrow\Delta+i\gamma/2$, where $\gamma$ is the spontaneous
emission frequency. However a large detuning is considered in the paper to avoid the effects of the
spontaneous emission.

Using convenient units of length ($k^{-1}$), momentum ($\hbar k$) and frequency
($\omega_{\nu}=\hbar k^{2}/2M$)), we can rewrite the Schr\"odinger equation in a dimensionless form:
\begin{equation}
\frac{d^{2}}{d\tilde z^{2}}\psi(\tilde z)=-\left \{\tilde q^{2}+\beta\tilde
z+\frac{1}{2}\tilde\Delta\pm\frac{1}{2}\sqrt{\tilde\Delta^{2}+4\tilde\Omega(\tilde z)^{2}}\right
\}\psi(\tilde z),
\label{Scheq2}
\end{equation}
in which the tilde denotes dimensionless units. In Eq.\ (\ref{Scheq2}), we find the function
$\tilde\Omega(z)=\tilde\Omega_{0}\exp\left(-\frac{(\tilde z-\tilde z_{c})^{2}\cos^{2}\phi}{\tilde
d^{2}}\right )\cos\left ((\tilde z-\tilde z_{c})\sin\phi\right )$. $\beta$ is a
gravitational parameter which introduces some differences depending on the mass of the atom. Since
we shall assume that $\tilde\Omega_{0}<<|\tilde\Delta |^{2}$, we can define a parameter
$\eta=\tilde\Omega_{0}^{2}/\tilde\Delta$ which determines the strength of the laser potential.
This can be easily shown by introducing a Taylor expansion on the right hand side of Eq.\
(\ref{Scheq2}):
\begin{equation}
\frac{d^{2}}{d\tilde z^{2}}\psi(\tilde z)=-\left\{\tilde q^{2}+\beta\tilde z-\tilde
V(z)\right\}\psi(\tilde z),
\label{Scheq3}
\end{equation}
where
\begin{equation}
\tilde V(\tilde z)=\eta\exp\left (-2\cos^{2}\phi (\tilde z-\tilde z_{c})^{2}/\tilde
d^{2}\right )\cos^{2}((\tilde z-\tilde z_{c})\sin\phi).
\label{Vz2}
\end{equation}
Note that the potential is formed by the product of a cosine squared function and a Gaussian
envelope.

In all the figures throughout the paper we analyse the case of $2s$--$2p$  $ ^{7}$Li
transition, whose parameters are $\lambda=670.8$ nm, $\omega_{\nu}=3.96\times 10^{5}$ s$^{-1}$,
$\gamma=3.72\times 10^{7}$ s$^{-1}$ and $\beta=2.93\times 10^{-4}$. The position of the center of
the laser region is at $z_{c}=300\sqrt{2}\pi k^{-1}=0.142$ mm, and the halfwidth of the laser
Gaussians is given by $d=100\pi k^{-1}=33.5$ $\mu$m.

\section{Spatial Gaps Image}
\label{sec:Spgaps}

In this section we present a model which allows us to understand the physics behind the results we
will observe in the following sections. From Eq. (\ref{Vz2}) we observe that the potential
$\tilde V(\tilde z)$ is quasi-periodic, except for the Gaussian envelope, its periodicity given by
$\Delta\tilde z=\pi/\sin\phi$. In order to better understand the effects of periodicity, let us
remove gravitation and spontaneous emission, and for the time being let us forget the smooth
Gaussian dependence. With these assumptions, the laser potential is a simple cosine
squared potential, whose amplitude ($\eta$) can be positive (if $\tilde\Delta>0$) or
negative (if $\tilde\Delta<0$), depending on which dressed state is reached. It is well known that a
periodic potential leads to an energy structure of allowed and forbidden bands
\cite{AshcroftMermin}. Only the incoming momentum components $q$ whose associated kinetic energy
lies within an allowed band can propagate inside the potential. If $q$ does not satisfy the band
condition (i.e. if
$q^{2}/2M$ lies in a gap), then the atoms with this incoming momentum cannot propagate inside the
laser region and are therefore reflected, reflection bands being formed. 

\begin{figure}[ht]
\begin{center}\
\psfig{file=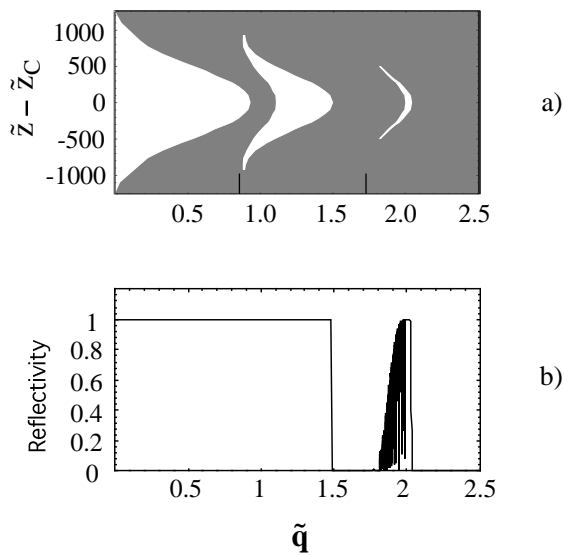,width=7.5cm}\\[0.3cm]
\caption{Results as a function of the $z$--momentum $q$ at $z=0$ for $\eta=2.0$, $\tilde
z_{C}=300\sqrt{2}\pi$, $\tilde d=100\pi$ and $\phi=\pi/3$. (a) Spatial band structure. White
regions correspond to three different spatial gaps. Small vertical bars indicate the
geometrical Bragg modes. The atoms are assumed to travel in the graph initially from down to
up, beginning with momentum $q$ at the bottom ($\tilde z=0$) of the figure. (b) Reflectivity
without gravitation and spontaneous emission.}
\label{fig:2}
\end{center}
\end{figure}

However, this image is excessively simple. In order to obtain physical insight into the
numerical results, obtained by direct resolution of the Schr\"odinger Eq.\
(\ref{Scheq3}), we must take into account the Gaussian laser envelope. The analysis is greatly
simplified if, as we consider in this paper, the width of the Gaussian envelope is very large
compared to the cosine squared periodicity. In particular, in the cases analysed below the width of
the Gaussian envelope of the potential at $1/e$ is $245$ times the cosine squared period. If this
condition is satisfied we can consider that within small intervals of the envelope we have a large
number of cosine squared oscillations of approximately constant amplitude. We can therefore define
an energy band structure for each of these intervals considered as the band structure calculated
for an infinitely extended periodic potential with this constant amplitude. We can extend this
reasoning and define a local band structure for each position
$z_{0}$ inside the laser region, or in other words for each value of the Gaussian envelope
($V_{env}(z_{0})$). This local band structure is calculated for a cosine squared potential of
infinite number of periods with constant amplitude $V_{env}(z_{0})$. Therefore each position
within the laser region is linked with an energetic spectrum of allowed and forbidden
energies, a spatially--dependent band--structure being formed. Fig.\ \ref{fig:2}a shows such
spatial band structure for the case of a repulsive potential of $\eta=2.0$. In particular, for
certain momenta some spatial regions are energetically forbidden (white regions). We will call
these regions {\em Spatial Gaps}. Fig.\ \ref{fig:2}b shows the reflectivity for different
incoming momentum components $q$, obtained by direct resolution of Eq.\ (\ref{Scheq3}). The spatial
gaps image allows an intuitive understanding of the physical processes behind these numerical
results. If an atom with some momentum component finds during its travel inside the laser region a
spatial gap, then the propagation through it is energetically forbidden, and consequently the
atom is reflected. This point becomes clear by comparison between Figs.\ \ref{fig:2}a and
\ref{fig:2}b. In \cite{Santos97} we also show that this model explains very well the effects of the
spontaneous emission and gravitation. In this paper we will assume a sufficiently large detuning to
neglect the spontaneous emission effects. The gravitational effects will be also negligible for
the momentum components of interest in this paper and for the laser arrangement considered.

\section{Fabry--P\'erot--like behavior}
\label{sec:Fabry}

In Sec.\ \ref{sec:Spgaps} we have observed that an atom is reflected if
it finds a spatial gap inside the laser region. Therefore the spatial gap acts as a
potential barrier. Note that :
\begin{itemize}
	\item If the spatial gap is sufficiently narrow then transmission via tunneling becomes
possible. This tunneling produced from one allowed spatial region to another is no more than
the well--known Landau--Zener tunneling between allowed energetic bands.
	\item  The form of the spatial gap depends on the specific incoming momentum one considers,
and hence this effective potential barrier has the interesting property that it changes its
form and width depending on the incoming atomic momentum.
\end{itemize}

\begin{figure}[ht]
\begin{center}\
\hspace{0mm}
\psfig{file=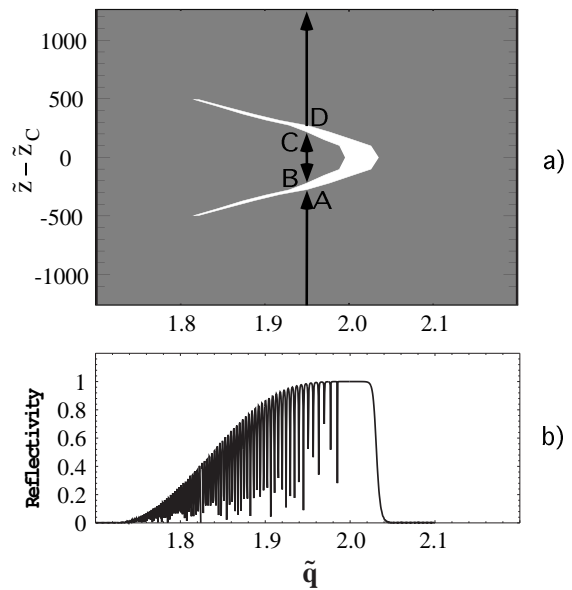,width=7.5cm}\\[0.3cm]
\caption{(a) Detail of Fig.\ \ref{fig:2}a. (b) Detail of Fig.\ \ref{fig:2}b. Note the appearance
of several resonances in the reflection spectrum.}
\label{fig:3}
\end{center}
\end{figure}

For certain interval of momenta (around $\tilde q =2$ in Fig.\ \ref{fig:2}b), several
peaks in the momentum spectrum of reflection can be observed. These peaks are depicted in
detail in Fig.\ \ref{fig:3}b. The physical process behind these peaks can be well
understood by using the spatial gaps image of Sec.\ \ref{sec:Spgaps}. In Fig.\ \ref{fig:3}a we
present in detail the region in Fig.\ \ref{fig:2}a corresponding with the region depicted in Fig.\
\ref{fig:3}b. As we observe in this figure the spatial gap, i.e. the effective potential
barrier, is very narrow in this region. This fact allows the possibility of tunneling of the
wavefunction through the spatial gap (from point $A$ to point $B$), and therefore a partial
reflection is produced. Since the envelope is symmetric, the transmitted part reaches again
an spatial gap (point $C$), from where the atoms can be partially transmitted again by
tunneling until point $D$, or can be reflected back to point $B$, from where they can be
partially reflected and so on. This process leads to multiple oscillations between points $B$
and $C$, i.e. the region from $B$ to $C$ acts as Fabry--P\'erot cavity \cite{BornWolf}, where $B$
and $C$ act as lossy mirrors. As we indicated previously, since the form of the effective
barrier depends on the momentum, the length of the "cavity" depends also on the momentum.
Following the optical analogy, this is the same case we would have if the length of a
Fabry--P\'erot cavity was different depending on the wavelength of the light. The peaks we
find in the momentum spectrum of the reflection can therefore be explained as resonances of
this special Fabry--P\'erot cavity.

\section{Bloch Oscillations and Wannier--Stark Ladders}
\label{sec:Bows}

\begin{figure}[ht]
\begin{center}\
\hspace{0mm}
\psfig{file=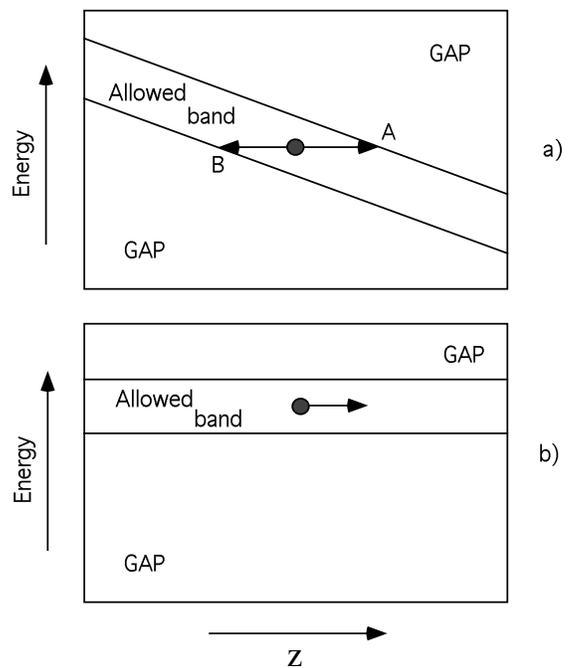,width=7.5cm}\\[0.3cm]
\caption{(a) Scheme of an atom evolving inside an allowed region between two forbidden ones
when an external constant force is applied. Therefore the atom is multiply reflected leading to
the well--known Bloch Oscillations. (b) Without external force there is no tilting of the
spatial gaps and the atom evolves freely inside the allowed band. Therefore the oscillations
are not produced.}
\label{fig:4}
\end{center}
\end{figure}

Other way to understand these resonances in the reflection spectrum is provided by a
well--known phenomenum of Solid--State Physics, namely the Wannier--Stark (WS) Ladders
\cite{Wannier60}. This effect is simply a stationary counterpart (i.e. in the
frequency domain) of another effect in the time domain called Bloch Oscillations (BO). Let
us briefly review this concept. BO appear when particles within a periodic
potential are affected by a constant acceleration. We must point out that although this
effect was initially observed in Solid--State context
\cite{Mendez93}, several recent experiments \cite{Dahan96} have reported
the same phenomena in Atom Optics. BO can be easily understood
for weak potentials \cite{Peik97}: due to the acceleration, the momentum of the particles
increases linearly according to Newton's law until it reaches a critical value satisfying
the Bragg condition, then the atomic wave is reflected and its momentum is reversed. The
atom travels again under Newton's law until it reaches other Bragg condition and then it is
reflected again. Then, the BO can be understood as multiple reflections between two Bragg
reflections. For larger potentials the BO can be understood using the  spatial gaps image
(as pointed out previously, Bragg reflection is a particular case of the band--gap structures
for weak potentials). Let us consider an infinitely--extended periodic potential of constant
amplitude affected by an external force. Let us suppose that this force is linear in the
spatial coordinate. Hence the band structure varies in the space leading to tilted allowed
and forbidden spatial regions (Fig.\ \ref{fig:4}a). Let us suppose  a particle inside an
allowed region. This particle moves until it reaches a forbidden region (point $A$), from
where it is reflected. Then it travels again within the allowed region until it finds
another forbidden region (point $B$), from where it is reflected and so on, leading to
multiple oscillations. Note that in absence of external force, no tilting is present (Fig.\
\ref{fig:4}b) and hence there are not multiple oscillations. 

We observe that the oscillations reported in Sec.\ \ref{sec:Fabry} are certainly similar to
the BO. In particular, both are due to the same reason, i.e. multiple reflections between
forbidden regions. However several important differences can be pointed out:
\begin{itemize}
\item The physical process which produces the "cavity" is basically different: in
the BO is the tilting of the bands due to an external force, while in our case is the
symmetry of the Gaussian envelope.
\item Whereas in the BO the reflection is produced between two
different spatial gaps, in the oscillations reported here both sides of the "cavity" are
produced by the same spatial gap, which is curved due to the Gaussian shape of the
laser envelope.
\item In the BO the atom is initially inside the allowed region between two forbidden
ones. In our case the atom enters from outside, and therefore needs to tunnel a narrow
spatial gap to enter into the internal allowed region where it oscillates.
\end{itemize}
Due to these similitudes and differences, we call the resonances in the reflection spectrum
Wannier--Stark--like resonances.

\section{Time Domain. Bloch--Like oscillations}
\label{sec:Bloch}

We have therefore interpreted the resonances appearing in Fig.\ \ref{fig:3}b as
Wannier--Stark--like resonances. As for standard WS--ladders, the Wannier--Stark--like
resonances are linked in the time domain with what we call Bloch--like oscillations. In order
to observe this effect we have numerically calculated the evolution of an atomic wavepacket
through the laser arrangement. In order to achieve this we have solved the Schr\"odinger Eq.\
(\ref{Scheq3}) evaluating the wavefuction inside and outside the laser region for different
incoming momentum components $q$. Note that at this point turns very important the possibility to
calculate the wavefunction inside the laser region, which cannot be calculated by using a
transfer--matrix method  as that of \cite{Tan94}. We have employed a finite--differencing numerical
method previously presented in \cite{Santos97}, which allows us to know the wavefunction also inside
the laser region. Once we have calculated the spatial behavior of the wavefunction for the different
momentum components $\psi (q,z)$, we obtain the temporal evolution of an atomic wavepacket
by applying a one--dimensional Fourier transform:
\begin{equation}
\phi (z,t)=\int e^{-i\frac{\hbar q^{2}t}{2M}}\psi (q,z) f_{0}(q) dq,
\label{FourierT}
\end{equation}
where $f_{0}(q)$ is an initial Gaussian momentum distribution of the form:
\begin{equation}
f_{0}(q)=e^{-a^{2}(q-q_{0})^{2}/2}e^{-iqz_{0}},
\label{inidistri}
\end{equation}
where $q_{0}$ is the central momentum component of the atomic wavepacket.

\begin{figure}[ht]
\begin{center}\
\hspace{0mm}
\psfig{file=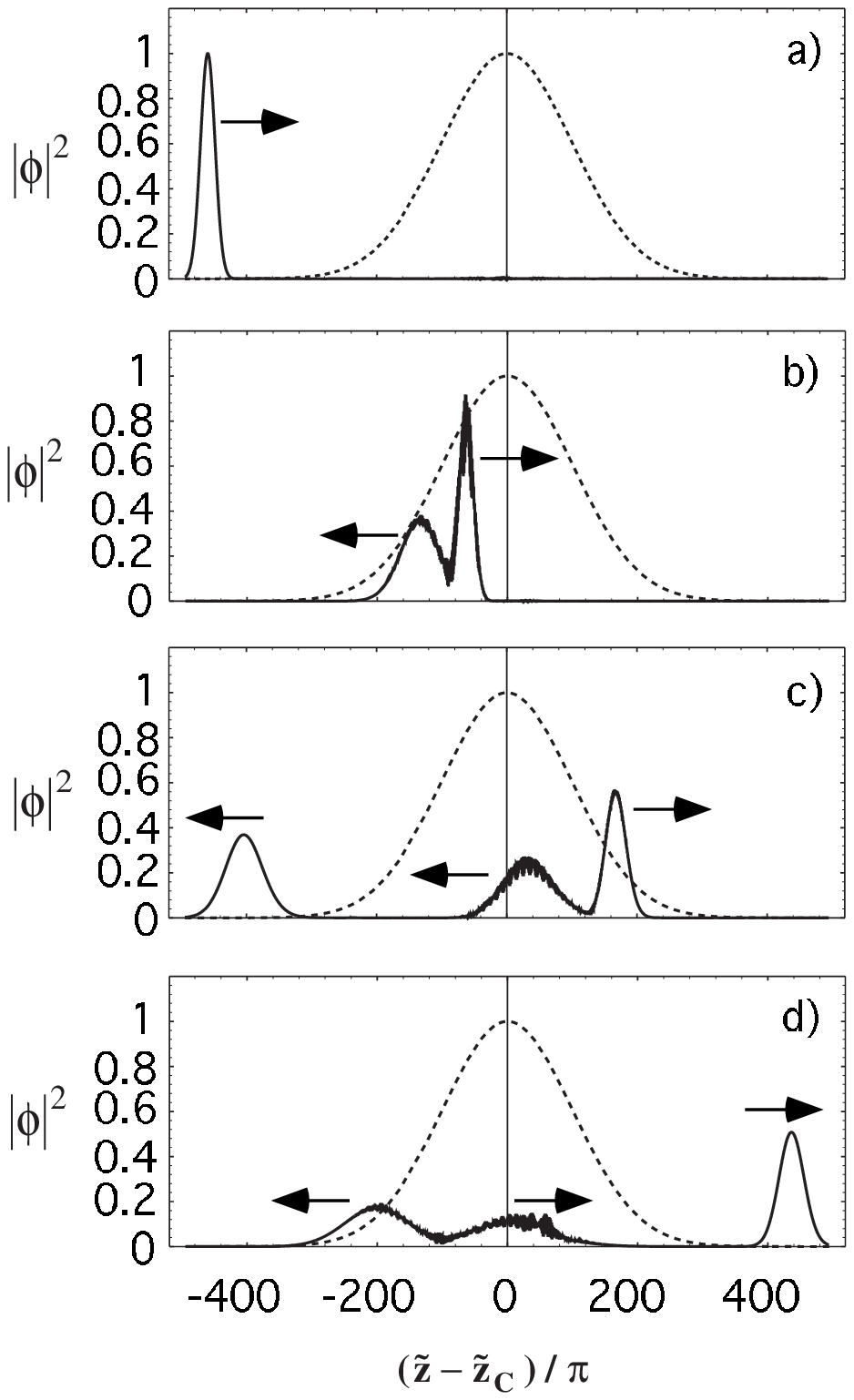,width=7.5cm}\\[0.3cm]
\caption{Temporal evolution of an atomic wavepacket through a laser region whose parameters
are those of Fig.\ \ref{fig:2}. The parameters of the wavepacket are $\tilde a=35\pi$,
$\tilde q_{0}=1.9$. (a) shows the density of probability $|\phi (z) |^{2}$ at time $t=0$, while
(b), (c) and (d) show respectively successive snapshots at $t=6T$, $t=10T$ and $t=14T$ where
$T=18\pi\omega_{\nu}^{-1}$. In all the figures the Gaussian shape of the laser envelope is shown in
dashed lines for comparative purposes (an arbitrary scale is used to depict the Gaussian).
The appearance of multiple oscillations inside the laser region is evident in this case.}
\label{fig:5}
\end{center}
\end{figure}

Fig.\ \ref{fig:5} shows the evolution of an atomic wavepacket with $\tilde a=35 \pi$,
$\tilde q_{0}=1.9$ through a laser arrangement as that of Fig.\ \ref{fig:3}. The initial position
of the centre of the atomic wavepacket is $\tilde z_{0}=-35\pi$. Fig.\ \ref{fig:5}a shows the
snapshot at $t=0$, whereas Figs.\ \ref{fig:5}b, c and d show respectively the snapshots at $t=6 T$,
$10 T$ and $14 T$, with $T=18\pi\omega_{\nu}^{-1}$. In all the figures the Gaussian form of the
laser envelope is represented in dashed lines for comparative purposes. In Fig.\ \ref{fig:5}b a
partial reflection and tunneling is clearly produced. The reflected wavepacket travels to the left
dissapearing, while the transmitted wavepacket travels within the laser region. In Fig.\
\ref{fig:5}c the previously transmitted wavepacket is again splitted into two parts: a transmitted
part which travels to the right dissapearing, and a reflected part which travels back in the laser
region. Again in Fig.\ \ref{fig:5}d a new splitting is produced. The observation of
successive multiple reflections becomes more difficult due to the wavepacket spreading and due
to the losses in the different partial transmissions.

\section{Quantum Multiple Reflections}
\label{sec:Qmr}

\begin{figure}[ht]
\begin{center}\
\hspace{0mm}
\psfig{file=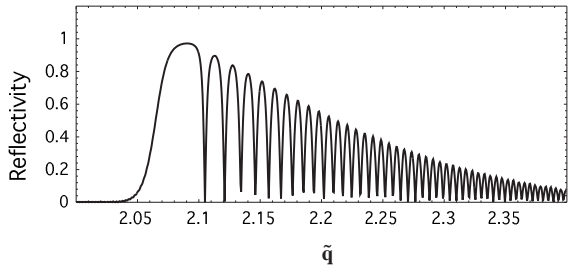,width=7.5cm}\\[0.3cm]
\caption{Reflectivity as a function of the incoming $z$--momentum $q$ at $z=0$ for $\eta=-5.0$,
$\tilde z_{C}=300\sqrt{2}\pi$, $\tilde d=100\pi$, and $\phi=\pi/3$. Note the appearance of several
resonances in the reflection spectrum.}
\label{fig:6}
\end{center}
\end{figure}

\begin{figure}[ht]
\begin{center}\
\hspace{0mm}
\psfig{file=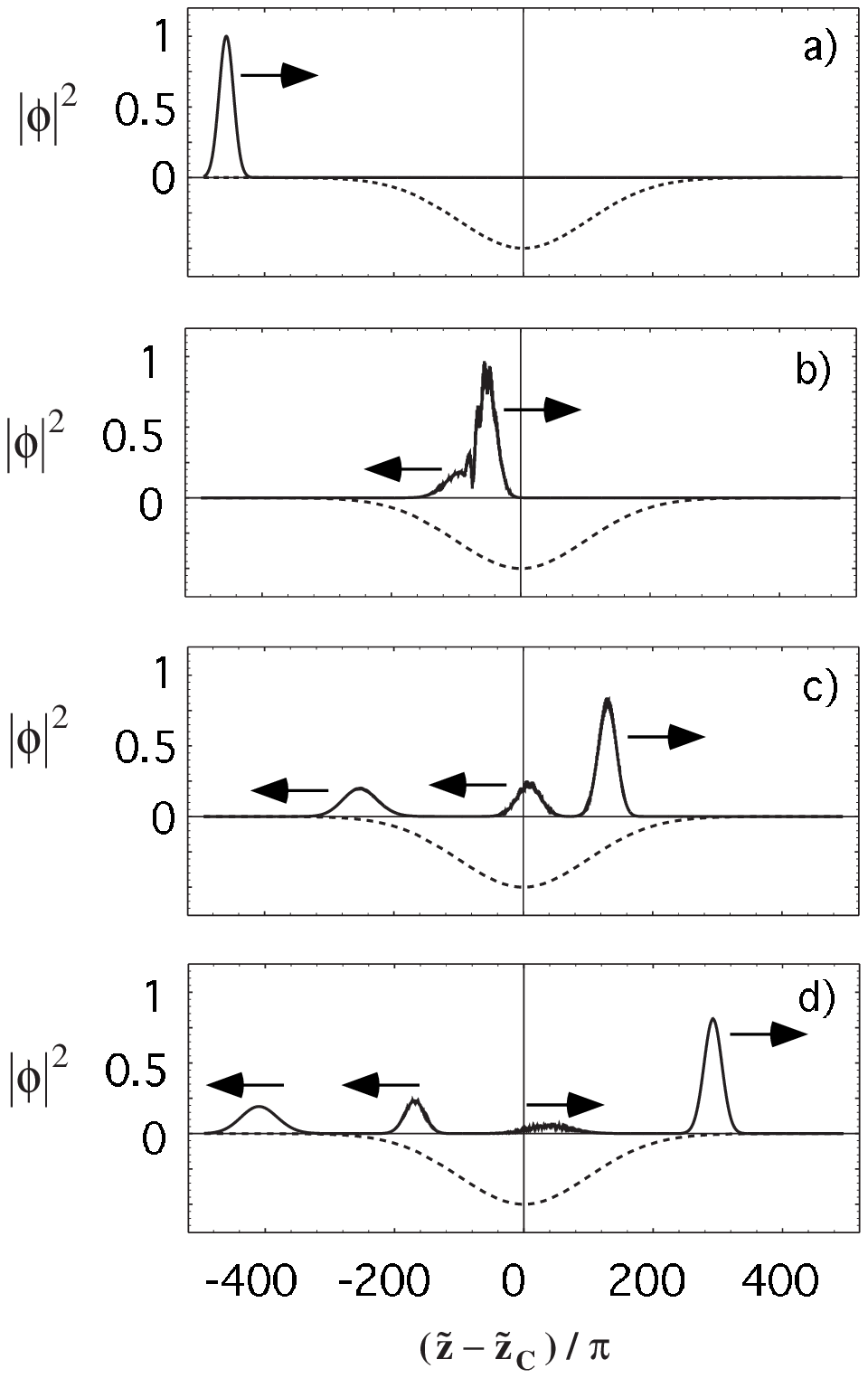,width=7.5cm}\\[0.3cm]
\caption{Temporal evolution of an atomic wavepacket through a laser region whose parameters
are those of Fig.\ \ref{fig:6}. The parameters of the wavepacket are $\tilde a=35\pi$,
$\tilde q_{0}=2.2$. (a) shows the density of probability $|\phi (z) |^{2}$ at time $t=0$, while
(b), (c) and (d) show respectively successive snapshots at $t=5T$, $t=7T$ and $t=9T$ where
$T=18\pi\omega_{\nu}^{-1}$. In all the figures the (negative) Gaussian shape of the laser envelope
is shown in dashed lines for comparative purposes (an arbitrary scale is used to depict the
Gaussian). The appearance of multiple oscillations inside the laser region is evident, even
considering that in this case the potential is attractive.}
\label{fig:7}
\end{center}
\end{figure}

We show in this section that the previously reported multiple oscillations can also
be observed  for the case of attractive laser potential, i.e. it is possible to achieve
multiple reflections in a laser potential that classically allows none. Fig.\ \ref{fig:6} shows a
detail of the reflection spectrum for the case of an attractive potential with $\eta=-5.0$. We
clearly observe the appearance of several resonances as those of Fig.\ \ref{fig:3}. These
distortions are no more than the previously analysed WS--like resonances. As for the repulsive
case, a narrow spatial gap appears leading to the already analysed effect of partial
Landau--Zener tunneling, and therefore to multiple reflections, whose effect in the momentum domain
is the appearance of the WS--like resonances. As in the previous case, Bloch--like oscillations can
be observed if we monitorize the evolution of an atomic wavepacket through the laser region. Fig.\
\ref{fig:7} shows this evolution for the case of the same atomic wavepacket of Fig.\
\ref{fig:5}, but now the laser is that of Fig.\
\ref{fig:6}. The first snapshot (Fig.\ \ref{fig:7}a) is at
$t=0$, whereas Figs.\ \ref{fig:7}b, c, d are respectively obtained at $t=5T$, $t=7T$,
$t=9T$, with $T=18\pi\omega_{\nu}^{-1}$, where we observe different wavepacket splittings due to
partial tunneling and reflection. As in the repulsive case multiple oscillations are possible, i.e.
quantum multiple reflections also appear.

\section{Conclusions}
\label{sec:Conclu}

In this paper we have analysed the reflection of an atomic beam dropped onto a laser with a
periodic profile modulated by a Gaussian envelope, formed in the interference region of two
Gaussian laser beams. We have numerically calculated the atomic reflection on such arrangement by
direct resolution of the corresponding Schr\"odinger equation, and explained the band--like
character of the reflection momentum spectrum using a band--theory model of energetically forbidden
spatial regions, or spatial gaps. We have probed that due to a partial Landau--Zener tunneling
through sufficiently narrow spatial gaps, and also due to the Gaussian symmetry of the laser
potential, an atom can undergo multiple reflections inside the laser region, in a process which
resembles a Fabry--P\'erot interferometer but for atomic--waves. These multiple oscillations have
been compared with the well-known Bloch Oscillations of Solid--State physics, analysing the
similarities and differences between both oscillations. In particular although both processes are
due to multiple reflections between energetically forbidden spatial regions, in the effect
presented in this paper the spatial gap is curved due to the Gaussian symmetry and the atom
oscillates between two barriers which are actually part of the same spatial gap, contrary to the
Bloch Oscillations in which the atom oscillates between two different spatial gaps. Also, the
reported multiple oscillations are not due to an external force as Bloch Oscillations but
due to the Gaussian form of the envelope. Due to these similarities and differences we have called
these oscillations Bloch--like oscillations. We have analysed the effect both
in the frequency domain (where Wannier--Stark--like resonances appear) and in the time--domain
(observing the Bloch--like oscillations). We have proved that the effect appears not only for
repulsive laser potentials, but also for attractive ones, i.e. quantum multiple reflections are also
possible.

\bigskip

{\bf Acknowledgments.} Partial support from the Spanish Direcci\'on General de Investigaci\'on
Cient\'\i fica y t\'ecnica (Grant No. PB95--0955) and from the Junta de Castilla y Le\'on
(Grant No. SA 16/98) is acknowledged.


\end{document}